\documentclass[aps,prb,preprint,superscriptaddress]{revtex4}
\usepackage{graphicx,bm}
\usepackage{color}
\usepackage{booktabs}

\newcommand{\EF}{\ensuremath{E_{\rm F}}}
\newcommand{\kB}{\ensuremath{k_{\rm B}}}
\newcommand{\kF}{\ensuremath{k_{\rm F}}}
\newcommand{\vF}{\ensuremath{v_{\rm F}}}
\newcommand{\rxx}{\ensuremath{\rho_{xx}}}
\newcommand{\rcor}{\ensuremath{\rho_{\rm osc}/\rho_{\rm BG}}}
\newcommand{\RH}{\ensuremath{R_{\rm H}}}
\newcommand{\nH}{\ensuremath{n_{\rm H}}}
\newcommand{\nQ}{\ensuremath{n_{\rm Q}}}
\newcommand{\Sxx}{\ensuremath{S_{xx}}}
\newcommand{\Sab}{\ensuremath{|S_{xx}|}}
\newcommand{\kxx}{\ensuremath{\kappa_{xx}}}
\newcommand{\nL}{\ensuremath{n_{\rm L}}}

\newcommand{\CdAs}{Cd$_{3}$As$_{2}$}
\newcommand{\ZnAs}{Zn$_{3}$As$_{2}$}
\newcommand{\CdZnAs}{Cd$_{3-x}$Zn$_{x}$As$_{2}$}

\begin{document}


\title{Doping-induced topological transition and enhancement of thermopower in the Dirac-semimetal system Cd$_{3-x}$Zn$_x$As$_2$ } 

\author{J.~Fujioka}
\email[corresponding author: ]{fujioka@ims.tsukuba.ac.jp}
\affiliation{Division of Materials Science, University of Tsukuba, 1-1-1 Tennodai, Tsukuba, Ibaraki 305-8573, Japan}
\author{M.~Kriener}
\email[corresponding author: ]{markus.kriener@riken.jp}
\affiliation{RIKEN Center for Emergent Matter Science (CEMS), Wako 351-0198, Japan}
\author{D.~Hashizume}
\affiliation{RIKEN Center for Emergent Matter Science (CEMS), Wako 351-0198, Japan}
\author{Y.~Yamasaki}
\affiliation{National Institute of Material Science, Tsukuba, Ibaraki, 305-0047, Japan}
\author{Y.~Taguchi}
\affiliation{RIKEN Center for Emergent Matter Science (CEMS), Wako 351-0198, Japan}
\author{Y.~Tokura}
\affiliation{RIKEN Center for Emergent Matter Science (CEMS), Wako 351-0198, Japan}
\affiliation{Department of Applied Physics and Quantum-Phase Electronics Center (QPEC), University of Tokyo, Tokyo 113-8656, Japan}
\affiliation{Tokyo College, University of Tokyo, Hongo, Tokyo 113-8656, Japan}

\date{\today}
\begin{abstract}
\CdAs\ is one of the prototypical topological Dirac semimetals. Here, we manipulate the band inversion responsible for the emergence of Dirac nodes by alloying \CdAs\ with topologically trivial \ZnAs. We observe the expected topological phase transition around a Zn concentration of $x\sim 1$ while the carrier density monotonically decreases as $x$ is increased. For larger $x$, the thermoelectric figure of merit exhibits comparably large values exceeding 0.3 at room temperature, due to the combined effects of a strong enhancement of the thermopower, an only moderate increase of the resistivity, and a suppression of the thermal conductivity. Complementary quantum-oscillation data and optical-conductivity measurements allow to infer that the enhanced thermoelectric performance is due to a flattening of the band structure in the higher-$x$ region in \CdZnAs.
\end{abstract}


\maketitle

\section{Introduction}
Quantum transport of relativistic electrons in topological semimetals has been an issue of great interest in topological materials' science \cite{armitage18a}. 
In such materials, the quantum state of the Dirac or Weyl electrons is strongly coupled to the crystal symmetry, 
and hence the engineering of the electronic symmetry is a promising way to search for exotic quantum transport of such quasiparticles. In recent years more and more materials have been theoretically predicted and  experimentally found to be a topological semimetal. Prototypical materials include $A_3$Bi with $A = $~Na, K, Rb \cite{zwang12b,zkliu14b,neupane14a}, BiO$_2$ and SbO$_2$ \cite{young12a}, and \CdAs\ \cite{zwang13a,ali14a,tliang14a,zkliu14a,sjeon14a,uchida17a,nakazawa18a,uchida19a}. 

Among them, \CdAs\ possesses a simple band structure with an electron charge carrier concentration of $\sim 10^{18}$~cm$^{-3}$. It has long been known for its large mobility of $\sim 10^{4}$~cm$^{2}$/Vs at room temperature \cite{turner61a}. 
Recently, an even higher value of almost $\sim 10^{7}$~cm$^{2}$/Vs was reported at low temperatures due to a linear band dispersion and strongly suppressed backscattering events of the charge carriers \cite{tliang14a}. 
The nontrivial topology of this system, namely, an inversion of conduction and valence bands which are of different character, manifests in two Dirac nodes in the proximity of the Fermi energy \EF\ \cite{zwang13a}, which are protected by both time-reversal symmetry and rotational symmetry of the crystal lattice \cite{zwang13a}. For example, it has been demonstrated that the breakdown of time-reversal symmetry via the application of a magnetic field creates a Weyl semimetalic state with negative magnetoresistance due to the chiral anomaly \cite{czli15a,jcao15a,hli16a,zjia16a}, a hallmark of the underlying nontrivial physics. 

Another way to control Dirac nodes in such systems is to manipulate the band inversion directly. 
It has been proposed that 
the chemical substitution of Cd with Zn changes the sign of band gap from negative (band inversion) to positive, 
resulting in the topological transition from a Dirac semimetal to a trivial insulator \cite{zdanowicz64a,zdanowicz64b,zdanowicz75a,hlu17a}. 
Indeed, in contrast to \CdAs, \ZnAs\ is a topologically trivial semiconductor with a hole carrier concentration of $\sim 10^{17}$~cm$^{-3}$ 
and a much lower room-temperature mobility of only $\sim 10$~cm$^2$/Vs \cite{turner61a}. 
Hence, a depletion of the charge carriers and a topological phase transition from the Dirac semimetal \CdAs\ to trivial \ZnAs\ is expected when alloying these two systems. Indeed, Lu {\it et. al,} found experimental indications of this transition in \CdZnAs\ in magnetotransport measurements \cite{hlu17a} around $x\sim 1.1$ on the basis of an enhanced resistivity upon cooling as well as a thorough analysis of Shubnikov-de-Haas (SdH) oscillations. Recent studies on thin films of \CdZnAs\ also support this scenario qualitatively \cite{nishihaya18a,nishihaya18b,nishihaya19a}, although the topological phase transition takes place already around $x\sim 0.6$ \cite{nishihaya19a}. We note that a similar transition is proposed to occur in related Cd$_3$As$_{2-x}$P$_x$ on the basis of angle-resolved photoemission spectroscopy data \cite{thirupathaiah18a}.

Given the remarkably high mobility of the electron charge carriers, \CdAs\ is expected to bear potential for a good thermoelectric performance 
with possibly large power factors $\Sxx^2/\rxx$; \Sxx\ and \rxx ~being the longitudinal thermopower and resistivity, respectively \cite{pei11a}. 
Indeed, a recent study reported $\Sxx^2/\rxx \sim 1.6 \times 10^{-3}$~W/K$^2$/m along with a fairly small thermal conductivity $\kxx \sim 5$~W/K/m, 
yielding $ZT \sim 0.1$ at room temperature \cite{czhang16a}; ZT represents the figure of merit $ZT= \Sxx^2 T/(\rxx\kxx)$ as a measure of the thermoelectric efficiency. 
This value further increases in presence of a magnetic field $B$, exceeding unity at $B = 7$~T and $T=375$~K \cite{hwang18a} mainly due to field-induced suppression of \kxx. 
Since these parameters also depend on the actual charge carrier concentration \cite{tzhou16a}, it is promising to study the thermoelectric performance upon doping.

In this study, we have measured transport, thermoelectric properties, and the charge dynamics upon Zn doping in Cd$_{3-x}$Zn$_x$As$_2$ with $0\le x \le 1.2$. 
With increasing $x$, the carrier density monotonically reduces and the Seebeck coefficient is largely enhanced, exceeding 300 $\mu$V/K at 300 K for $x=1.2$. 
At low temperatures, we could confirm the reported metal-insulator transition with Zn doping \cite{hlu17a}. At the same time, Zn doping suppresses the thermal conductivity while the resistivity above the metal-insulator transition temperature is enhanced only modestly due to the doping-induced disorder. Hence, the thermoelectric figure of merit is greatly enhanced, exceeding 0.3 at room temperature, i.e., more than three times the value reported for pure \CdAs. Complementary analyses of quantum oscillation and optical conductivity data suggest an $x$-dependent change in the band-structure dispersion in the higher doping region which promotes the enhancement of the figure of merit.

This paper is organized as follows: First, we will present electric and thermal transport data with enhanced $ZT$ values. Then we will analyze magnetotransport and optical spectroscopy data which point toward the scenario of an $x$-dependent change of the band structure at \EF\ giving rise to the observed large room-temperature $ZT$ values. We will finish with a discussion of our findings and conclude with a summary of the paper.

\section{Experimental Methods}
Single-crystalline samples of Cd$_{3}$As$_2$ were grown by the Bridgmann technique, while polycrystalline samples of Cd$_{3-x}$Zn$_x$As$_2$ were synthesized by conventional melt-growth. In both cases, stoichiometric ratios of the constituent elements were mixed inside a glove box, transferred into quartz tubes, and eventually sealed while evacuated. 
In the Bridgman-method growth, 
the temperature of the upper (lower) heater was set to 900$^{\circ}$C (600$^{\circ}$C). 
The evacuated quartz tubes were kept for 12~h at 900$^{\circ}$C and then lowered with a speed of 2~mm/h. 
After the quartz tubes had reached the lower heater, they were slowly cooled down to room temperature. 
Melt-grown batches were kept for 48~h at 800$^{\circ}$C -- 950$^{\circ}$C depending on the composition and slowly cooled down to room temperature afterwards.

Resistivity and Hall effect were measured by a conventional five-probe method in a commercial system (PPMS, physical property measurement system, Quantum Design). The thermopower and thermal conductivity were measured in a home-built setup inserted into a PPMS while applying a temperature gradient by using a chip-heater attached on one side of the sample. The temperature gradient is monitored by employing commercial thermocouples. The reflectivity spectra at nearly normal incidence were measured between room temperature and 10~K in the energy region of 0.008 -- 5~eV. In the case of single-crystalline \CdAs, a sample surface with $[1 1 \bar{2}]$-orientation was polished. Then the spectra were measured with $[1\bar{1}0]$ light polarization. As for \CdZnAs, reflectivity spectra were measured with unpolarized light. A Fourier transform spectrometer and a grating-type monochromator equipped with a microscope were employed in the photon energy range 0.008 -- 0.7~eV and 0.5 -- 5~eV, respectively. Measurements in the energy range of 3 -- 40~eV were carried out at room temperature by using synchrotron radiation at UV-SOR, Institute for Molecular Science (Okazaki). For Kramers-Kronig transformations, we adopted suitable extrapolation procedures for energy ranges which were not accessible by the used experimental setups: below 0.008~eV the Hagen-Rubens-type (metal) or constant-reflectivity (insulator) extrapolation was used, respectively. Above 40~eV an $\omega^{-4}$-type extrapolation was utilized.


\section{Results}
\begin{figure}[htbp!]
\centering
\includegraphics[width=0.7\linewidth]{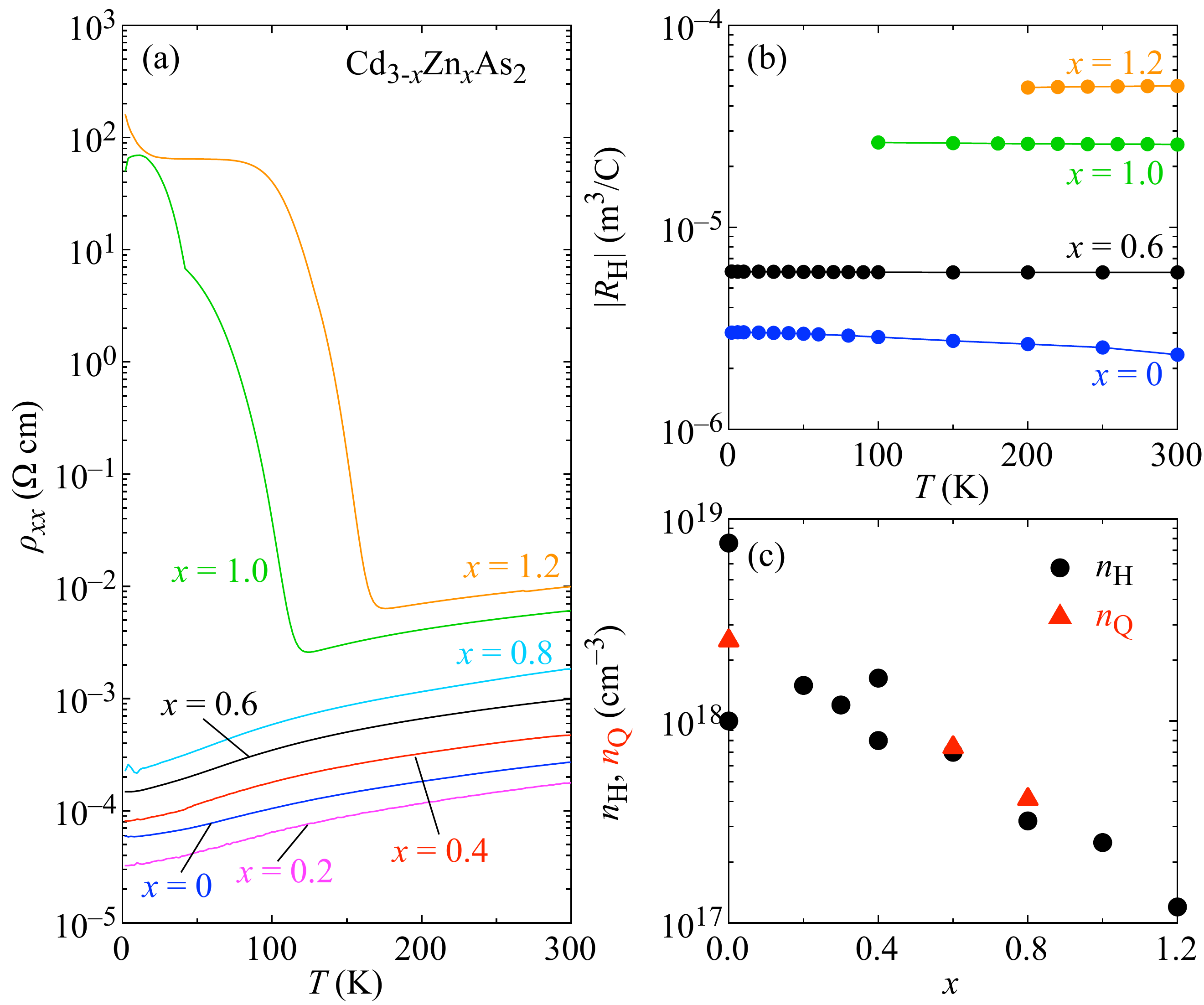}
\caption{Temperature dependence of (a) resistivity and (b) absolute value of the Hall coefficient $|\RH|$ for \CdZnAs. Here, \RH\ is negative for all $x$ and the whole temperature range examined. (c) Carrier densities as a function of doping level $x$. Here, \nH\ (\nQ) denotes the carrier density derived from Hall coefficient data (quantum oscillations). 
\nQ is derived with assuming an isotropic Fermi surface. }
\label{fig1}
\end{figure}

Figure~\ref{fig1}(a) shows the temperature dependence of the longitudinal resistivity \rxx\ for \CdZnAs. In the low-doped region ($0 \le x \le 0.6$), the resistivity decreases upon lowering temperature, i.e., the system behaves like a metal. The residual resistivity at 5~K is enhanced with increasing $x$ as compared to our pure \CdAs\ sample except for $x=0.2$. An upturn is  clearly observed around $\sim 120$~K and $\sim 170$~K for $x=1.0$ and $x=1.2$, respectively, highlighting the metal-to-insulator transition in these higher-doped samples. The overall qualitative temperature dependence of the resistivities of $x = 1.0$ and 1.2 is similar. However, at very low temperatures there is a downturn in \rxx\ of the sample with $x=1.0$, while the resistivity of the sample with $x=1.2$ increases again after exhibiting a broad plateau between $\sim 30$~K and $\sim 80$~K. These features are clearly distinct from what is expected for a conventional insulator, the resistivity of which monotonically increases upon decreasing temperature.

Figure~\ref{fig1}(b) summarizes the temperature dependence of the absolute value of the Hall coefficient \RH. For all $x$, \RH\ is nearly temperature independent and its sign is negative, indicating that the conduction in all examined samples is of electron type. Estimated carrier densities \nH\ at room temperature assuming a single carrier model are plotted against respective Zn concentrations in Fig.~\ref{fig1}(c), together with \nQ\ estimated from quantum-oscillation data (see Fig.~\ref{fig3}). 
As expected the absolute value of the carrier density monotonically decreases as a function of $x$ from the order of a few times $10^{18}$cm$^{-3}$ for $x=0$ down to $1.2 \times 10^{17}$cm$^{-3}$ for $x=1.2$, reflecting the depletion of the electron-type carriers when going from $n$-type \CdAs\ to $p$-type \ZnAs. However, the charge neutrality point, i.e., the Cd:Zn ratio where the sign change of \RH\ takes place, is not reached up to $x=1.2$. 
We observe this crossover in slightly higher-doped samples around $x\sim 1.5$ (not shown). 
The metallic samples $x \leq 0.8$ investigated here exhibit mobilities of about $\sim 10^5$~cm$^2$/V/s at 2~K and $\sim 10^4$~cm$^2$/V/s at 300~K, respectively. 
We note that several properties, such as residual resistivity, charge carrier concentration etc.\ 
of this material are rather sample dependent as shown in Fig.~\ref{fig1}(c); 
for $x=0$ and 0.4, there are exemplarily shown two charge carrier concentrations measured on two different samples, respectively. 
Such and even larger variations have been also reported for the parent material \CdAs, see, e.g., Ref.~\onlinecite{tliang14a}. 
This is possibly related to differences in the (Cd,Zn):As ratio. In \CdAs, ideally one fourth of the Cd lattice sites are unoccupied and these vacancies seem to order in a chiral way along the $c$ axis which may differ from sample to sample even if these samples were cut from the same initial batch \cite{prvComm}, cf.\ also the discussions in Refs.~\onlinecite{ali14a} and \onlinecite{tliang14a}.

\begin{figure}[htbp!]
\centering
\includegraphics[width=0.7\linewidth]{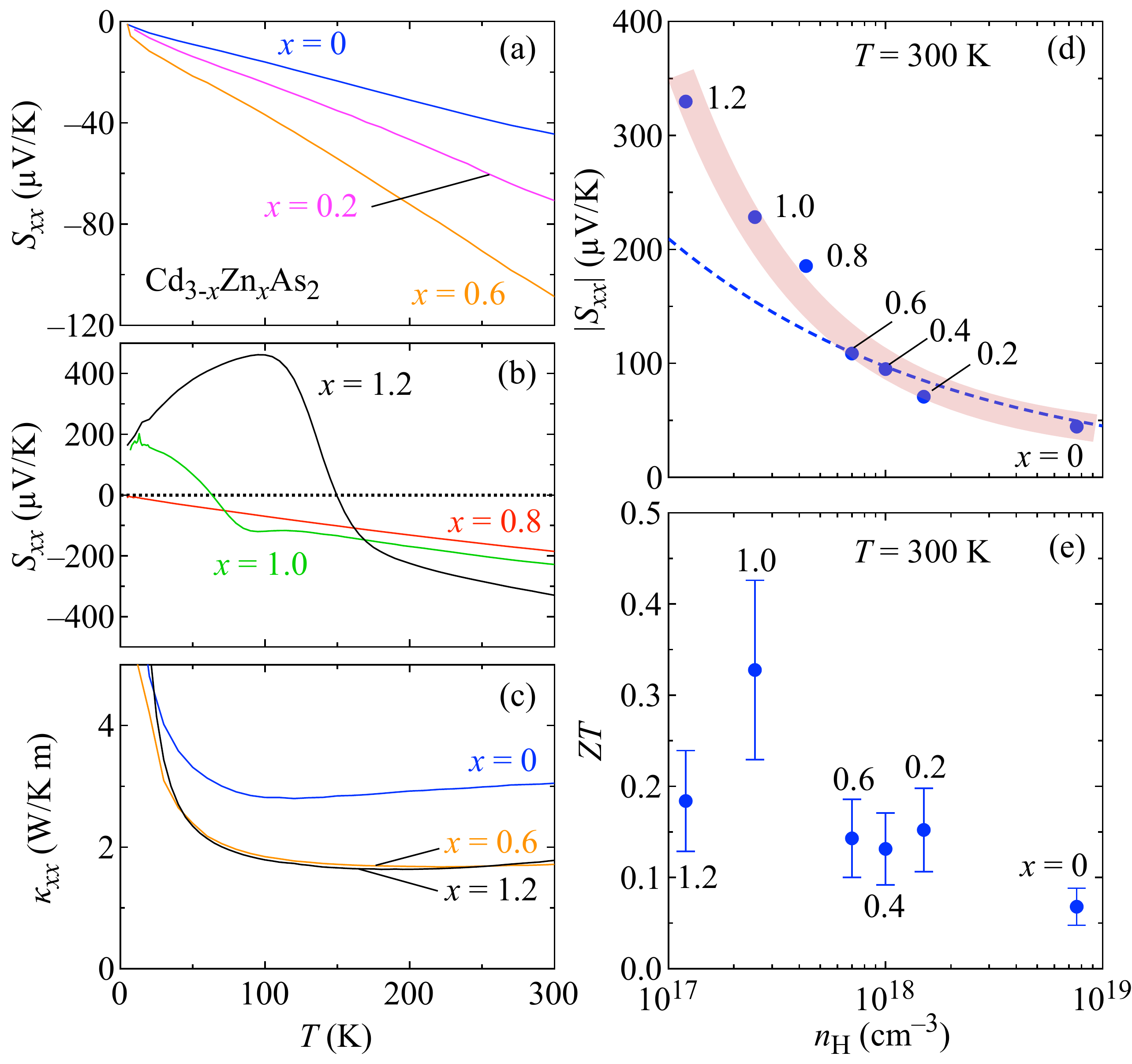}
\caption{Temperature dependence of the Seebeck coefficient (a) for $x = 0$, 0.2, and 0.6 and (b) for $x=0.8$, 1.0, and 1.2. (c) Thermal conductivity for  $x = 0$, 0.6, and 1.2. (d) Absolute value of the Seebeck coefficient $|\Sxx|$ and (e) figure of merit $ZT$ at 300~K plotted as function of the charge carrier concentration $\nH$ for $0 \leq x \leq 1.2$. In (d), the dashed line indicates the semiclassical expectation $\Sab\propto n^{-1/3}$ assuming a $k$-linear band dispersion. The broad lightly-red shaded curve is merely a guide to the eyes.}
\label{fig2}
\end{figure}

\begin{figure}[htbp!]
\centering
\includegraphics[width=0.7\linewidth]{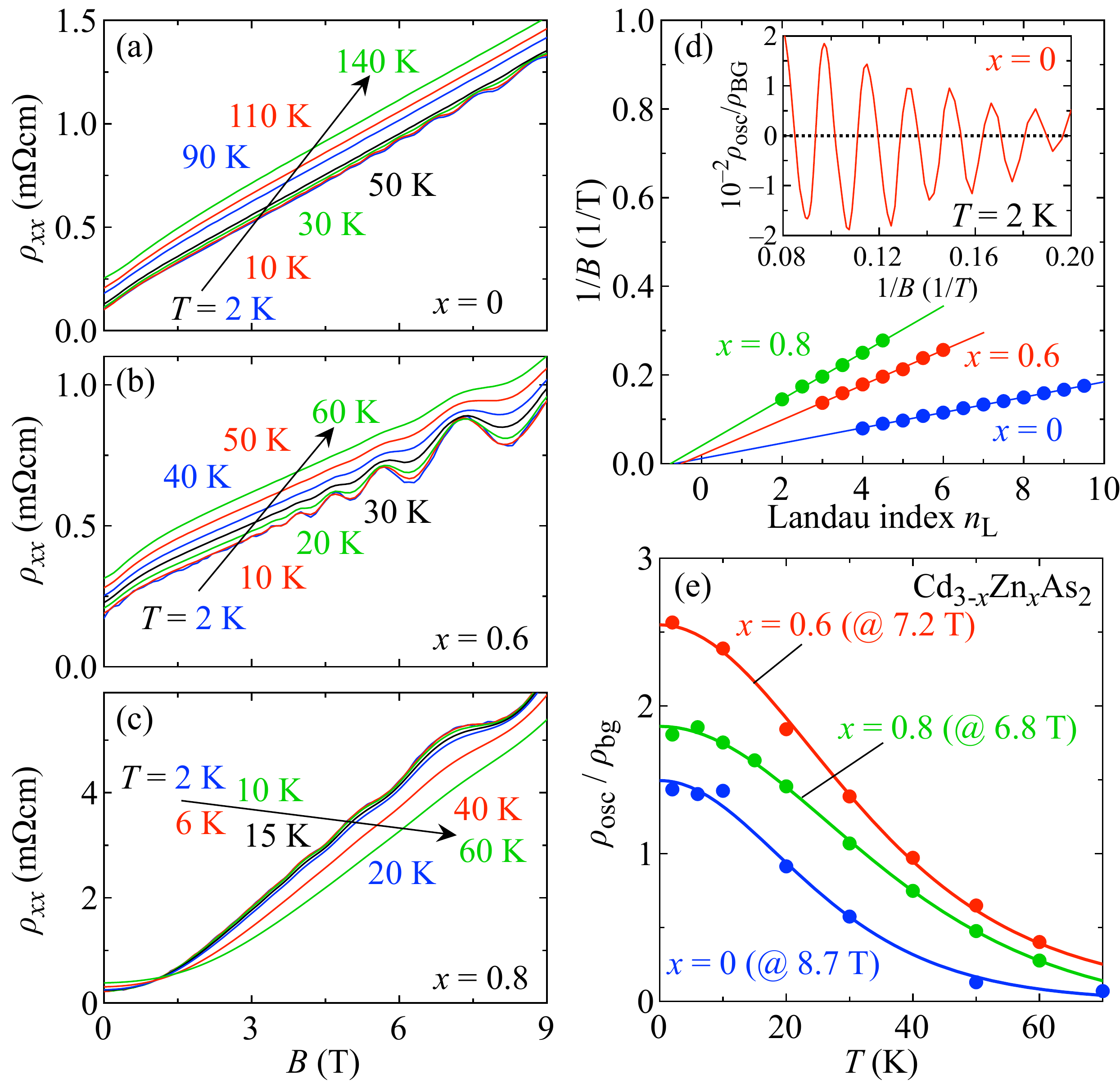}
\caption{Magnetoresistivity at various temperatures for (a) $x =0$, (b) 0.6 and (c) 0.8, respectively. (d) Landau level fan diagram for $x =0$ (blue), 0.6 (red), and 0.8 (green) at $T=2$~K. In the inset, background-corrected quantum oscillations \rcor\ for $x=0$ are shown as a function of $1/B$. (e) Temperature dependence of the amplitude of \rcor\ at selected field strengths as indicated in brackets next to the respective Zn concentrations [color code is the same as in panel (d)].}
\label{fig3}
\end{figure}

Thermoelectric and thermal-transport data are summarized in Figs.~\ref{fig2}. 
The temperature dependence of the Seebeck coefficient \Sxx\ is shown in Figs.~\ref{fig2}(a) and (b) for $x \leq 0.6$ and $x\geq 0.8$, respectively. 
In the lightly-doped region $x\leq 0.8$, \Sxx\ is negative and nearly proportional to temperature, which is often observed in conventional metals and semiconductors. 
By contrast, \Sxx\ exhibits a nonmonotonic temperature dependence for larger $x$: below approximately 100~K and 170~K, 
\Sxx\ deviates significantly from a temperature-linear behavior for $x=1.0$ and 1.2, respectively. 
In particular, \Sxx\ exhibits a sign change and becomes positive upon further cooling. 
Moreover, these temperatures nearly coincide with the upturn observed in resistivity data [see \ Fig.~\ref{fig1}(a)]. 

The longitudinal thermal conductivity \kxx\ is shown for selected $x$ in Fig.~\ref{fig2}(c). 
For all samples, \kxx\ is almost temperature independent down to $\sim 100$~K but steeply increases towards lower temperatures possibly due to an enhancement of the phonon mean-free path. Interestingly, in the thermal conductivity there are no characteristic anomalies visible between 50~K and 200~K in clear contrast to resistivity (steep upturn) and thermopower data (clear slope change) even for $x = 1.2$, where these are most pronounced.

Absolute values of the Seebeck coefficient \Sab\ at 300 K are replotted as a function of charge carrier concentration \nH\ in Fig.~\ref{fig2}(d). 
The respective Zn concentrations $x$ are given for each data point. Apparently, \Sab\ increases monotonically with decreasing \nH: For our pure \CdAs\ sample, we find $\Sab = 44~\mu$V/K. For $x=1.2$, \Sab\ is enhanced by more than a factor of six, exceeding 300~$\mu$V/K. 
This behavior is qualitatively consistent with the case of typical semiconductors or metals, where, according to Mott's formula, \Sab\ is inversely proportional to \EF, which decreases here as indicated by the depletion of the electron carrier concentration with $x$, cf.\ Fig.~\ref{fig1}(c). The dashed line in Fig.~\ref{fig2}(d) indicates the expected charge-carrier-concentration dependence of \Sab\ ($\propto n^{-1/3}$) in the semiclassical framework of Mott's formula with the assumption of a $k$-linear band dispersion. 
Apparently, this line fits well to the experimental data for $x \leq 0.6$ but clearly falls short for larger $x$.

The presented quantities allow us to calculate the thermoelectric figure of merit $ZT = \Sxx^2 T/(\rxx\kxx)$, 
the room-temperature values of which are plotted against \nH\ in Fig.~\ref{fig2}(e). 
As compared to pristine \CdAs\ ($ZT = 0.07$), $ZT$ increases with $x$ and exhibits a maximum $ZT= 0.33$ for $x=1.0$, 
a fairly large room-temperature value of the figure of merit. 
Here, we anticipate error bars of 30\% because the values of \rxx, \Sxx, and \kxx\ are not precisely reproducible and depend on the sample used for the measurement, as already discussed above.

To obtain further insight into what mechanism might be responsible for the observed enhancement of the thermoelectric efficiency as represented by $ZT$, we investigated the impact of Zn doping on the electronic structure in \CdZnAs\ by analyzing magnetoresistivity. 
Experimental data along with analyses of SdH oscillations are summarized in Fig.~\ref{fig3}. 
The magnetoresistivity for $x=0$, 0.6, and 0.8 are shown in Figs.~\ref{fig3}(a), (b), and (c), respectively. 
For $x=0$ and $0.6$, the resistivity is nearly proportional to the magnetic field 
and exhibits quantum oscillations, i.e. Shubunikov-de Haas (SdH). 
Such a $B$-linear magnetoresistivity is often observed in Dirac semimetals and is one characteristic feature of the highly mobile Dirac electrons \cite{armitage18a}. 
Similar SdH oscillations are also observed for $x=0.8$ while the magnetoresistivity is rather quadratic in $B$ in the low-field region.

Figure~\ref{fig3}(d) contains the corresponding Landau level (LL) fan diagrams with the oscillation frequency $1/B$ plotted against the Landau index \nL. 
These were extracted according to the Lifshitz–Onsager quantization rule $B_F/B = \nL - \phi$ from the data shown in panels (a) -- (c) after subtracting the background magnetoresistivity $\rho_{\rm BG}$ by approximating it with a polynomial. The resulting oscillation part \rcor\ is exemplarily shown for $x=0$ in the inset to Fig.~\ref{fig3}(d). Then we assigned integer and half-integer indexes to the peak and valley positions in the magnetoresistivity data, respectively, as described in more detail, e.g., in Ref.~\onlinecite{maryenko15a}. 
The linearity of the fan plot up to the quantum limit may be a consequence of small Zeeman splitting in this system. 
From the slope of the LL fan diagrams, the oscillation frequency $B_F$ is estimated to be 58~T, 25~T, and 18~T for $x=0$, 0.6, and 0.8, respectively. 

Figure~\ref{fig3}(e) shows the temperature dependence of the background-corrected quantum oscillations $\rho_{\rm osc}/\rho_{\rm BG}$ at selected magnetic fields. From the thermal damping of the oscillation amplitudes upon warming, the cyclotron mass is estimated to be $0.051 m_0$, $0.033 m_0$, and $0.029 m_0$ in units of the bare electron mass $m_0$ for $x=0$, 0.6, and 0.8, respectively, by employing the Lifshitz-Kosevich formula \cite{maryenko15a}. We note that the Fermi velocity is nearly independent of the carrier density, suggesting that the band dispersion is close to $k$ linear in this range of $x$. Table~\ref{tab1} summarizes these and additional parameters extracted from the SdH oscillations.

\begin{figure}[htbp!]
\centering
\includegraphics[width=0.9\linewidth]{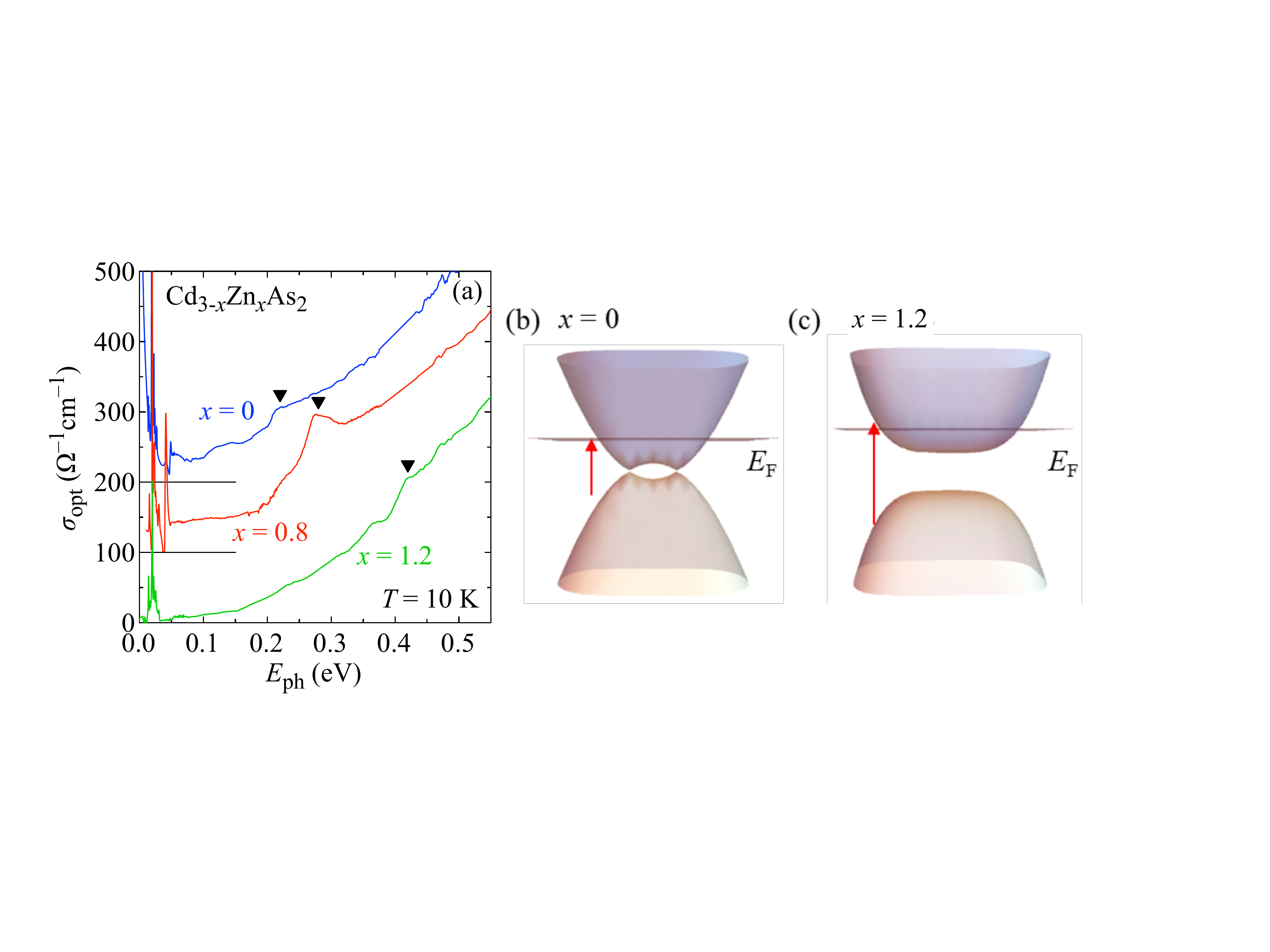}
\caption{(a) Optical conductivity spectra at 10~K for $x=0$ (blue), 0.8 (red), and 1.2 (green). The spectra are shifted by 100~$\Omega^{-1}$cm$^{-1}$ with respect to each other for clarity. Solid black lines indicate the respective offsets. Triangles denote the threshold of interband transitions. The sharp peaks below 0.1~eV are attributed to optical phonons. Schematic illustration of the band dispersion for (b) $x=0$ and (c) $x=1.2$. The red arrow denotes the optical interband transition at the threshold energy marked with triangles in (a). The horizontal plane indicates the Fermi level \EF. See text for details.}
\label{fig4}
\end{figure}

To obtain further insight into the electronic state, Fig.~\ref{fig4}(a) shows the optical conductivity spectra at 10 K for $x=0$, 0.8, and 1.2. 
Spiky structures below 0.1~eV are ascribed to phonon excitations. 
As a common feature in all the three samples, the interband electron excitation from the valence to the conduction band manifests itself 
as a very slow increase of the optical conductivity as a function of the phonon energy, 
which is often observed in gapless or small gap semimetals/semiconductors \cite{akrap16a,neubauer16a,crassee18a, Jenkins2016, Fujioka2021, Chen2015}. 
Moreover, for each sample a small peak or kink is observed at about 0.2, 0.3, and 0.4~eV for $x=0$, 0.8, and 1.2, respectively, 
as indicated with black triangles in Fig.~\ref{fig4}(a). 
We note that similar features are identified in the data taken at different temperatures, assuring that these kinks are an intrinsic feature. 
The kink is most remarkable in the case of $x=0.8$. 
Such an absorption peak/kink was often observed and interpreted as the threshold of the interband transition \cite{akrap16a,neubauer16a,crassee18a}. 
Apparently, this threshold energy is enhanced as $x$ is increased. 
Taking into account that the carrier density is monotonically reduced upon increasing $x$, it is likely that the topological transition has occurred and a gap has opened in the case of the larger Zn concentration $x=1.2$ as schematically illustrated in Fig.~\ref{fig4}(c), in comparison with $x = 0$ shown in Fig.~\ref{fig4}(b). 

\section{Discussion}
Finally we will discuss the relevance of the observed electronic structure to the observed enhancement of the figure of merit exceeding 0.3 at room temperature. 
In the present case of a Dirac dispersion, the Fermi energy should be scaled to the Fermi wave number \kF\ which is proportional to $n^{1/3}$. According to Mott's formula, \Sab\ is inversely proportional to \EF\ and, thus, is expected to scale with $n^{-1/3}$. 
As shown in Fig.~\ref{fig2}(d) (dashed curve), the charge-carrier-concentration dependence of \Sab\ is consistent with this semiclassical scaling for higher carrier densities, i.e., above $6\times 10^{17}$~cm$^{-3}$ which corresponds to $x \lesssim 0.6$. 
However, when further increasing the Zn concentration, the coincidence becomes worse and eventually deviates significantly when the electron carriers become very diluted. 
In general, quantum oscillations are a highly sensitive probe of the electronic states in the vicinity of \EF\ while the Seebeck coefficient is strongly influenced or determined by the electronic states in an energy range of $\pm 4\kB T$ around \EF \cite{Usui2017}. 
Hence, in the present case, the Seebeck coefficient may probe the energy dispersion in the energy range of $\EF/\kB\pm 1200$~K. Thus, the significant discrepancy of the experimental Seebeck coefficients and the expectation in the semiclassical model is likely to indicate that the band dispersion away from \EF\ is not linear in $k$ any more in the heavily Zn-doped samples with $x>0.6$ as sketched in Fig.~\ref{fig4}(c). This strongly supports our initial working hypothesis that Zn doping is an efficient tool to tailor and finely tune the band structure in the Dirac semimetal \CdAs\ and should eventually trigger the topological phase transition.

The remaining question to be addressed is the origin of the thermally induced metal-insulator transition 
as indicated by the pronounced enhancement of \rxx\ below $\sim 200$~K for $x \geq 0.8$, 
which is also reflected in the nonmonotonous temperature dependence of the thermopower. 
Older literature reported a doping-induced structural transition in \CdZnAs\ \cite{zdanowicz64b}. 
In order to look for a possible link between these two features, we performed temperature-dependent powder x-ray diffraction experiments on a sample with $x = 1.2$, 
but could not find any hint for a structural change upon cooling \cite{Supple}. 
Hence, the origin of this remarkable temperature-dependent change in resistivity and thermopower remains unclear and remains to be an interesting 
phenomenon to be elucidated for future studies.

\section{Summary}
In summary, we demonstrate a topological transition in the Dirac semimetal \CdAs\ by engineering the band structure by replacing Cd with its lighter counterpart Zn with weaker spin-orbit interaction. 
Associated with this transition, the bands at the Fermi level are flattened and a strong enhancement of the thermopower is successfully induced. 
Moreover, the thermal conductivity is suppressed while the resistivity remains reasonably small, 
yielding a fairly large figure of merit $ZT = 0.33$ at $T=300$~K. 
Our findings demonstrate that doping is an easy but highly efficient tool to control the topologically nontrivial band structure in Dirac semimetals and that such systems can be very promising starting points to look for an enhanced thermoelectric performance.

\section*{Acknoledgements}
\noindent
We thank D. Maryenko, T. Koretsune, R. Arita, T. Ideue and T. Liang for useful discussions and technical support. 
This work was partly supported by Grant-In-Aid for Science Research (Nos. 24224009, 15K05140, 16H00981, 18H01171, 18H04214, 16H06345) from the MEXT, 
and by PRESTO(No. JPMJPR15R5) and CREST(No. JPMJCR16F1), JST (No. JP16H00924), Japan. 
JF and MK contributed equally to this work.

%

\newpage

\newpage

\newpage

\newpage

\newpage
\begin{table*}[h]
\centering
\caption{Results of the analyses of the Shubnikov-de Haas (SdH) oscillations in magnetoresistivity data. Here, \kF\ denotes the Fermi wave vector, \nQ\ the electron-type charge carrier concentration as estimated from SdH oscillations, $m_{\rm c}/m_0$ gives the cyclotron mass in units of the electron mass $m_0$, \vF\ the Fermi velocity, $T_{\rm D}$ refers to the Dingle temperature, and $\tau_{\rm Q}$ to the respective scattering time.}
\vspace{0.5cm}
\label{tab1}
\begin{tabular}{ccccccc}
\toprule 
$x$  & \kF\ (\AA$^{-1}$) & \nQ\ (cm$^{-3}$) & $m_{\rm c}/m_0$ & \vF\ (m/s)         & $T_{\rm D}$ (K) & $\tau_{\rm Q}$ (s)   \\ \hline \addlinespace[0.75em]
 0   & 0.042             & $2.5\times 10^{18}$     & 0.051           & $9.4\times 10^{5}$ & 12              & $1.0\times 10^{-13}$ \\ \addlinespace[0.5em]  
 0.6 & 0.028             & $7.4\times 10^{17}$     & 0.033           & $9.9\times 10^{5}$ & 26              & $4.7\times 10^{-14}$ \\ \addlinespace[0.5em]    
 0.8 & 0.023             & $4.1\times 10^{17}$     & 0.029           & $9.3\times 10^{5}$ & 10              & $1.2\times 10^{-13}$ \\ \addlinespace[0.5em]  
\bottomrule
\end{tabular}
\end{table*}

\end{document}